\documentclass[iop]{emulateapj}
\usepackage{amsmath}
\usepackage{apjfonts}
\usepackage{gensymb}

\newcommand\xone{x_1}
\newcommand\xtwo{x_2}

\newcommand\cs{c_s}

\newcommand\Msun{\; {\rm M}_{\odot}}
\newcommand\Lsun{\; {\rm L}_{\odot}}
\newcommand\kms{\; {\rm km}\;{\rm s}^{-1}}
\newcommand\pc{\;{\rm pc}}

\newcommand\kpc{\;{\rm kpc}}
\newcommand\kpcre{\;{\rm kpc}^{-1}}
\newcommand\gravunit{\kpc \Msun^{-1}\;{\rm km}^2\;{\rm s}^{-2}}
\newcommand\freq{\kms\kpc^{-1}}

\newcommand\Myr{\;{\rm Myr}}

\newcommand\Surf{\Msun\;{\rm pc^{-2}}}

\newcommand\RCR{R_{\rm CR}}

\newcommand\Omb{\Omega_{b}}

\newcommand\simgt{\lower.5ex\hbox{$\; \buildrel > \over \sim \;$}}
\newcommand\simlt{\lower.5ex\hbox{$\; \buildrel < \over \sim \;$}}

\def\lvplot{($l,v$) diagram}
\def\lvplots{($l,v$) diagrams}

\def\spose#1{\hbox to 0pt{#1\hss}}
\def\dt{\spose{\raise 1.0ex\hbox{\hskip2pt$\mathchar"201$}}}    

\shorttitle{Gas dynamics in the Milky Way}

\shortauthors{Li et al.}

\begin{document}

\title{Gas dynamics in the Milky Way: a low pattern speed model} 

\author{Zhi Li\altaffilmark{1,2,3,4}, Ortwin Gerhard\altaffilmark{3,5}, Juntai Shen\altaffilmark{1,6}, Matthieu Portail\altaffilmark{3,7} and Christopher Wegg\altaffilmark{3,8}} 

\affil{$^1$Key Laboratory for Research in Galaxies and Cosmology, Shanghai Astronomical Observatory, Chinese Academy of Sciences, 80 Nandan Road, Shanghai 200030, China \\
$^2$University of China Academy of Sciences, 19 A Yuquanlu, Beijing 100049, China \\
$^3$Max-Planck-Institut f\"{u}r Extraterrestrische Physik, Giessenbachstrasse, 85748 Garching, Germany \\ 
Emails:$^4$lizh@shao.ac.cn $^5$gerhard@mpe.mpg.de $^6$jshen@shao.ac.cn $^7$portail@mpe.mpg.de $^8$wegg@mpe.mpg.de}



\begin{abstract}

We present gas flow models for the Milky Way based on high-resolution grid-based hydrodynamical simulations. The basic galactic potential we use is from a N-body model constrained by the density of red clump giants in the Galactic bulge. We augment this potential with a nuclear bulge, two pairs of spiral arms and additional mass at the bar end to represent the long bar component. With this combined model we can reproduce many features in the observed \lvplot\ with a bar pattern speed of $33\freq$ and a spiral pattern speed of $23\freq$. The shape and kinematics of the nuclear ring, Bania's Clump 2, the Connecting arm, the Near and Far 3-kpc arms, the Molecular Ring, and the spiral arm tangent points in our simulations are comparable to those in the observations. Our results imply that a low pattern speed model for the bar in our Milky Way reproduces the observations for a suitable Galactic potential. Our best model gives a better match to the \lvplot\ than previous high pattern speed hydrodynamical simulations.

\end{abstract}


\section{Introduction}

Studies of atomic and molecular gas in the inner Galaxy have revealed strong non-circular motions, which are now understood to be caused mainly by the Galactic bar. Because distances to individual gas clouds are difficult to obtain, these data are commonly presented in \lvplots, which show the distribution of gas emission line intensity as a function of Galactic longitude and line-of-sight velocity. Streamers and arms in the gas flow appear as high density lines in the \lvplot\ and, because of the unknown distances, must be interpreted through gas dynamical models.

Hydrodynamic models of the gas flow in the Milky Way have been able to reproduce many of the distinctive features in the \lvplots\ for HI and CO data \citep{bur_lis_93,dame_etal_01}, even though no model has been able to provide a good match to all the observed features \citep{eng_ger_99,fux_99b,rod_com_08,baba_etal_10,sorman_etal_15c}. A variety of barred potentials were used, including potentials derived from COBE or star count data, or potentials characteristic of barred N-body models, Besides the bar, also the Galactic spiral arms play some role for the gas flow \citep{bissan_etal_03,seo_kim_14}, by regulating the inflow of gas into the bar region.

Apart from the gravitational potential, the pattern speed of the bar is the most important parameter for the gas flow, because for given mass distribution it sets the resonance radii where the gas flow needs to accommodate the transition from one closed orbit family to another. A number of early investigations concluded a relatively high value for the pattern speed, $50-65\freq$ \citep{eng_ger_99,fux_99b,debatt_etal_02,bissan_etal_03}, such that the co-rotation radius of the bar would be located in the range $\RCR=3.5-5$ kpc, but others have argued for lower values \citep{wei_sel_99,rod_com_08,shen_14,sorman_etal_15c}.

Recently \citet{weg_ger_13} measured the three dimensional density of red clump giants (RCG) in the barred Galactic bulge. This density, together with the kinematics from the BRAVA survey \citep[]{kunder_etal_12}, has since been used as a constraint in constructing dynamical models of the barred bulge using the made-to-measure method \citep{portai_etal_15a}. Surprisingly, these models required a rather low pattern speed for the bar, $25-30\freq$, so that $\RCR>7$ kpc.

In these models the bulge represents the central buckled, box/peanut part of a longer bar similar to many N-body bars \citep[]{com_san_81, raha_etal_91, martin_etal_06,shen_etal_10}. Star count studies extending to larger longitudes have indeed found a thinner bar outside the Milky Way's barred bulge \citep[e.g.][]{hammer_etal_00, benjam_etal_05, cabrer_etal_08,wegg_etal_15} that ends near $l \approx 27-30\degree$. If these two components are aligned and form a single structure at an angle to the Sun of $\sim 27\degree$, as suggested by \citet{mar_ger_11} and found with the detailed RCG maps of \citet{wegg_etal_15} then the `long bar' component ends at $>4.7\kpc$ from the Galactic Center. Because the bar cannot exist beyond co-rotation \citep{contop_80} this limits the pattern speed to $\Omb\leq47\freq$ for a flat rotation curve at $220\kms$, lower than found by the majority of previous gas dynamics studies, but still significantly larger than the best-fitting pattern speed in the barred bulge dynamical models from \citet{portai_etal_15a}.

The purpose of the present paper is to enquire whether the observed \lvplot\ can be explained by a gas flow model in such a low $\Omb$ model, if we use a potential based on these dynamical models as an input to study the gas flow in a realistic Milky Way context. 

The paper is organized as follows: In Section 2, we describe our galaxy models, model parameters, and the numerical method. In Section 3, we present the best-fitting gas model, and explore the parameter space in Section 4. In Section 5, we discuss our assumptions and implications for our model, and summarize our results. 

\begin{figure}[!t]
\epsscale{1.0} \plotone{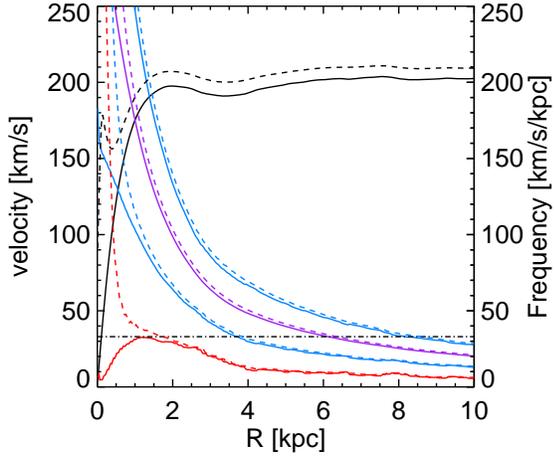}
\caption{Rotational velocity of gas in the potential of P15 (solid lines) and of our best-fitting model (dashed lines), together with the respective sets of frequency curves. Red lines denote $\Omega-\kappa/2$, outer and inner blue lines denote $\Omega \pm \kappa/4$, and the purple line denotes $\Omega$, respectively. Here $\Omega^2\equiv R^{-1}d\Phi_{\rm gal}/dR$ and $\kappa^2\equiv R^{-3}d(R^4\Omega^2)/dR$ denote the angular and epicyclic frequencies (\citealt{bin_tre_08}). The horizontal dot-dashed line represents our bar pattern speed ($33\freq$). The differences between these two rotation curves are caused by adding the nuclear bulge, the spiral arms, and the long bar.
\label{fig:rotcurve}}
\vspace{0.2cm}
\end{figure}

\section{Galaxy Models And Numerical Method} 

\subsection{Hydrodynamical Simulation}

We study here how gas responds to an imposed barred-spiral galaxy potential ($\Phi_{\rm gal}$) by using hydrodynamic simulations. The potential is described in Section 2.2. The bar and the spiral are assumed to rotate rigidly about the Galaxy Center with two different fixed pattern speeds. We solve the following dynamical equations of ideal gas in the frame co-rotating with the bar in $z=0$ plane: 
\begin{equation}\label{eq:con}
\left(\frac{\partial}{\partial t}  + \mathbf{u}\cdot\nabla \right) \Sigma
= - \Sigma\nabla\cdot\mathbf{u},
\end{equation}
\begin{equation}\label{eq:mom}
\left(\frac{\partial}{\partial t}  + \mathbf{u}\cdot\nabla \right) \mathbf{u}
= -\cs^2 \frac{\nabla \Sigma}{\Sigma} - \nabla \Phi_{\rm gal} + \Omb^2
\mathbf{R} - 2\mathbf{\Omb}\times \mathbf{u}.
\end{equation}
Here $\Sigma$, $\mathbf{u}$, and $\mathbf{\Omb}$ denote the gas surface density, the gas velocity, and the bar pattern speed, respectively.

Equation \eqref{eq:con} and \eqref{eq:mom} are solved by a modified version of the grid-based MHD code \textit {Athena} \citep{gar_sto_05,stone_etal_08,sto_gar_09}. By adopting a higher-order Godunov scheme, \textit {Athena} conserves mass and momentum of the fluid within machine precision. We integrate the equations on a uniform Cartesian grid with $2048\times2048$ cells which corresponds to a square box with a length of $L=30\kpc$. The grid spacing is therefore $\Delta x=\Delta y=14.6\pc$, comparable to the size of giant molecular clouds (GMCs). These high-resolution runs are necessary to capture the instabilities and turbulence of gas (also see \citealt{sorman_etal_15a}). We take the van Leer algorithm with piecewise linear reconstruction, choose the first-order flux correction, adopt the \textit{exact} Riemann nonlinear solver, and apply the outflow boundary conditions at the domain boundaries (i.e., at $|x|=L/2$ or $|y|=L/2$) for our hydrodynamic models. We do not impose a point symmetry relative to the Galaxy Center, thus allowing odd-$m$ modes to grow in our models.  

\begin{figure*}[!t]
\epsscale{1.2} \plotone{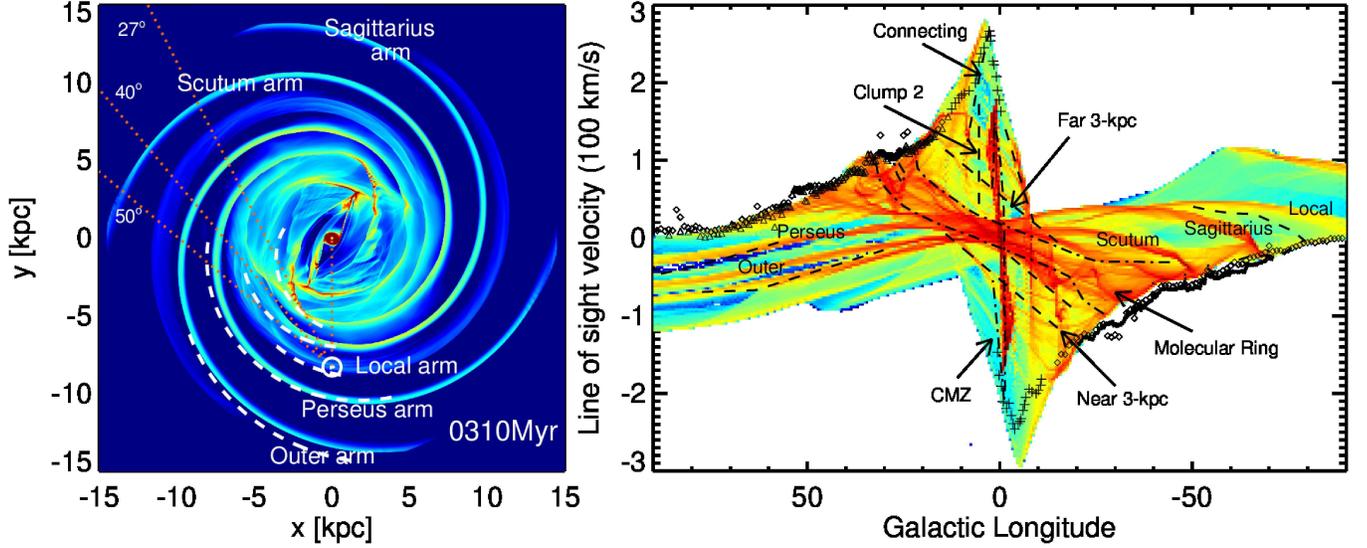}
\caption{This figure illustrates the gas surface density (left panel) and the corresponding \lvplot\ (right panel) of our best-fitting model. The snapshot is taken at a time when the spiral arms have completed a whole rotation relative to the bar ($310\Myr$). Left panel: Red and blue colors show gas with high and low densities. The bar major axis has a angle of $27\degree$ respect to the Sun-Galactic Center line. The solar symbol denotes the position of the Sun at ($0\kpc$, $-8.3\kpc$), from which the four red dotted lines represent four galactic longitude directions ($0\degree$, $27\degree$, $40\degree$ and $50\degree$). We also label the names of the major spiral arms. The white dashed lines are the locations of spiral arms from \citet{reid_etal_14} determined from HMSFRs. Right panel: the dashed and dotted-dashed lines represent various features on the \lvplot\ from \citet{rod_com_08}. The crosses show HI terminal velocities from \citet{mcc_dic_07}, the diamonds those from \citet{fich_etal_89}, and the plus signs those from \citet{bur_lis_93}. The triangles show CO terminal velocities from \citet{clemen_85}.
\label{fig:besfit}}
\vspace{0.2cm}
\end{figure*}

The real interstellar medium (ISM) in the Milky Way is multiphase and turbulent. The gas temperatures may differ by as much as a few orders of magnitude from inner region to outer part of the galaxy (e.g. \citealt{field_etal_69,mck_ost_77,mck_ost_07}). We model the gaseous disk to be isothermal, two-dimensional, unmagnetized, and non-self-gravitating. The thin disk approximation for the Milky Way is appropriate as the gas layer is very thin ($\sim100\pc$) for the most of the disk region, and that the large-scale ($l,v$) features are influenced primarily by the underlying large-scale Galactic potential is also reasonable. The limitations of the simplified gas description are summerized in \citet{li_etal_15}. In this paper we fix the effective isothermal sound speed $\cs$ to be $10\kms$ which is the same as previous studies (e.g. \citealt{fux_99b,rod_com_08}), and set up an exponential gas disk with surface density:

\begin{equation}\label{eq:surfden}
\Sigma_{\rm gas}(R) = \Sigma_{\rm 0}{\rm exp}(-R/R_{\rm gas}).
\end{equation}

The coefficient $\Sigma_{\rm 0}$ and the scale length of the gas disk $R_{\rm gas}$ here are $130\Surf$ and $3.5\kpc$ respectively, which gives a total gas disk mass of $1.0 \times 10^{9}\Msun$.

\subsection{Gravitational Potential}

Our gravitational potential is constructed from the superposition of several components. We began from a barred N-body potential, but found that a nuclear bulge, two pairs of spiral arms and an enhanced thin long bar were needed to produce reasonable gas flow models. These potential components are observationally constrained and added for physical reasons with preselected functional forms. Therefore we cannot expect to obtain a perfect \lvplot\ that fits observations in all details. The influences of each of these components on the gas flow is described in Section ~\ref{sec:simulationvariation}.

\subsubsection{N-body Potential}

The basis of the galacic potential is the dynamical model M80 constructed by \citet[][hereafter P15]{portai_etal_15a}. This model was obtained by fitting the 3D density of red clump giants (\citealt{weg_ger_13}) to match the BRAVA kinematics (\citealt{kunder_etal_12}) with the made-to-measure (M2M) method \citep[][P15]{delore_etal_07}. We take the mid-plane potential from this 3D M2M N-body model so that the box-peanut shape of the bulge is taken into account. The bar angle to the line of sight is $(27\pm2)\degree$ and the axis ratios of the bar are ($10:6.3:2.6$). Because the data cover the inner $10\degree$ of the Galaxy we expect the potential to be accurate in this region. However, as the gas flow depends on the full large-scale potential, this model requires modification. Therefore we include additional components in the next subsections.

\subsubsection{Nuclear Component}

The N-body model was constrained by off-plane data which does not extend to the Galactic Center. In the central $\lesssim 300\pc$ observations show the presence of an additional nuclear bulge component \citep{launha_etal_02}, which is not faithfully represented in the N-body model since it is too close to the Galactic plane and not extended enough along the line-of-sight to be resolved by the RCG star counts. We therefore add a potential calculated from the model density found by \citet{launha_etal_02}:

\begin{equation}\label{eq:nb}
\Phi_{\rm nb}(R) = -\frac{\gamma_{\odot}}{2}LG(1+nR\Gamma(0.2,c)-q\Gamma(0.4,c))/R.
\end{equation}

According to \citet{launha_etal_02}, the nuclear bulge is a superposition of two components, both of which follow the equation above. $G=4.302\times10^{-6}\gravunit$ is the gravitational constant, $\Gamma$ is the incomplete gamma function, $c=(pR)^{5}$, and $L$, $n$, $p$, $q$ are coefficients. We use a mass-to-light ratio $\gamma_{\odot}$ of $2$ suggested by \citet{launha_etal_02}. For the first component, $L=8.09 \times 10^{8}\Lsun$, $n=3.48\kpcre$, $p=7.74\kpcre$ and $q=0.45$; for the second component these values are $L=4.88 \times 10^{8}\Lsun$, $n=1.90 \kpcre$, $p=4.22\kpcre$ and $q=0.45$, respectively. The total mass of the nuclear bulge component is $1.4 \times 10^{9}\Msun$. Such a compact nuclear bulge is necessary for making reasonable gas flow patterns, especially for the Central Molecular Zone (CMZ), as we show later in the paper.

\begin{figure*}
\epsscale{1.2} \plotone{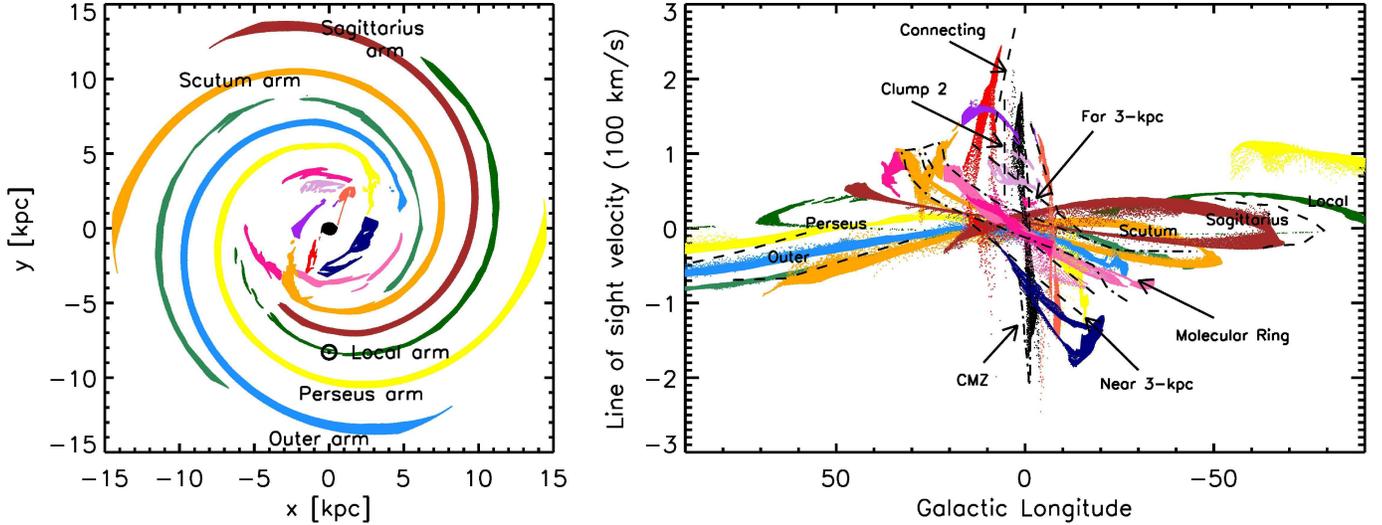} 
\caption{Link between the structures in the $x-y$ plane and the features in the $l-v$ plane in our best-fitting model using the same color coding. 
\label{fig:features}}
\end{figure*}

\subsubsection{Spiral Arms}

\label{sec:sppotential}

The gas under a spiral arm perturbation would show clear density contrast between the arm and inter-arm regions, which produces features and voids in the \lvplot\ \citep{bissan_etal_03}. Using over 100 trigonometric parallaxes of masers associated with young high-mass stars, \citet{reid_etal_14} mapped out the detailed shapes of the four major spiral arms in the Milky Way. For simplicity we include the four spiral arms with two separate, but identical in shape, pairs of logarithmic $m=2$ spiral arm potentials, offset by $20.25\degree$ in azimuth. We parametrize the potential of each pair of the $m=2$ spiral arms with a form motivated by \citet{junque_etal_13}:

\begin{eqnarray}\label{eq:spiral}
\Phi_{\rm sp}(R,\varphi) =
\left\{ \begin{array}{ll}
-\zeta_{\rm sp}R&e^{-\frac{R^2}{\sigma^2}[1-\cos(m\varphi-f_{\rm
m}(R))-R/\epsilon_{\rm sp}]}  ~~ R \geq R_{\rm sp} \\
-\zeta_{\rm sp}R&e^{-\frac{R^2}{\sigma^2}[1-\cos(m\varphi-f_{\rm
m}(R))-R/\epsilon_{\rm sp}]}  \\
~~~~ \times&e^{-(R-R_{\rm sp})^2/2\sigma_{\rm sp}^2}
 ~~~~~~~~~~~~~~~~~~~~ R < R_{\rm sp},
\end{array} \right.
\end{eqnarray}

with the shape function as 
\begin{equation}\label{eq:spiral_core} 
f_{m}(R) = \frac{m}{\tan(i)}{\rm ln}(R/R_{\rm i})+\gamma.
\end{equation}

Here $m=2$ representing two spiral arms, $\zeta_{\rm sp}=-800 \,(\rm km/s)^2$ is the perturbation amplitude, $\epsilon_{\rm sp}=4\;{\rm kpc}$ is the scale length of the spiral, $i=12.5^\circ$ is the pitch angle, $\sigma=2.35\;{\rm kpc}$ is the half-width of the spiral arms in the azimuthal direction (the true width in a direction perpendicular to the arms is given by $\sigma_{\perp}=\sigma \sin i$), $R_i=8\kpc$, and $\gamma=139.5^\circ$ and $69.75^\circ$ for these two pairs of arms are just the phase angles. Inside of $R_{\rm sp}=9\kpc$, the spiral strength decays towards the center in a Gaussian form with a dispersion $\sigma_{\rm sp}$=$1.5\kpc$. This is used for abating the effects of spiral arms inside the bar's co-rotation radius \citep{kim_ost_06}. The Local arm forms self-consistently in our simulations. All the spiral arms in our best-fitting model match to the shape of observed spiral arms \citep{reid_etal_14} reasonably well (see Section \ref{sec:simulationresults}).

In order to give a rough estimate of the mass of the spiral arms we used, we assume a vertical isothermal sheet potential profile for the spiral arm model in Equation \eqref{eq:iso}:

\begin{equation}\label{eq:iso}
\begin{aligned}
\Phi_{\rm sp}(R,\varphi,z) = \Phi_{\rm sp}(R,\varphi) &\times& \left(\frac{z_{0}}{1\kpc}{\rm ln}[{\rm cosh}(z/z_{0})]-1 \right)\\
&\propto& \frac{z_{0}}{1\kpc}{\rm ln}[{\rm cosh}(z/z_{0})]-1,
\end{aligned} 
\end{equation}

which approximately corresponds to a 1D vertical density profile $\rho_{\rm sp}(z) \propto (1\kpc/z_{0}){\rm sech}^{2}(z/z_{0})$ for $z_0<<R$. Then we use the 3D Poisson equation $\nabla^{2}[-\Phi_{\rm sp}(R,\varphi,z)] = 4\pi G\rho_{\rm sp}(R,\varphi,z)$ to derive the volume density and integrated mass. This does not exactly correspond to the analytic profile, and may have some negative densities which however have a small effect as long as $z_{0}$ is less than $0.2\kpc$. For our mass estimate we adopt $z_{0}=0.05\kpc$  as suggested by \citet{wegg_etal_15}. The resulting total mass of the spiral arms is then $2.1\times10^{9}\Msun$ within $2z_{0}$ $(0.1\kpc)$, and varies only slightly if we use $z_{0}=0.1\kpc$.

The pattern speed of the spiral arms should be lower than the bar pattern speed, so that the spirals could have a larger co-rotation radius which is important for channeling a continuous gas flow inwards (Section~\ref{sec:effectsofspirals}). In a recent work, \citet{junque_etal_15} used open clusters and the red giants from APOGEE to derive a robust estimate of the spiral pattern speed to be $23.0 \pm 0.5\freq$. We tested other spiral pattern speed values and find that the best values to give a reasonable gas flow are within the range of $21-24\freq$. Therefore we adopt $\Omega_{\rm sp} = 23.0\freq$. This value is consistent with previous simulations (e.g. \citealt{bissan_etal_03,pettit_etal_14}).

\subsubsection{Long Bar and Leading Ends}

\label{sec:lepotential}

\begin{figure*}
\epsscale{1.0} \plotone{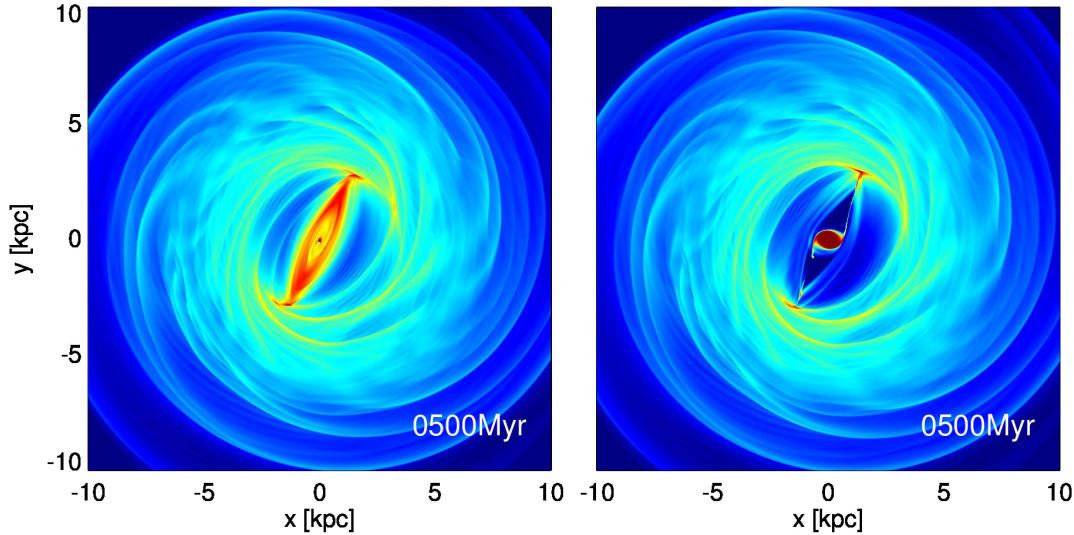} 
\caption{The effects of the nuclear bulge. The Left panel shows the $x_1$-type ring in the gas flow of the original P15 model, and the right panel shows the $x_2$-type ring in the P15 model plus a nuclear bulge. The snapshots are taken at $t=500\Myr$ at which the gas flow reaches quasi-steady state. The spiral arms here are excited by the rotating bar potential, and they weaken at the bar co-rotation radius ($6.3\kpc$).
\label{fig:nb}}
\end{figure*}

The N-body model was based on star counts that were not constrained outside the bulge, so that the mass of the the bar outside the bulge, or long bar, is underestimated. The existence of a thin long bar component in the Milky Way is indicated by a variety of NIR and MIR star counts (\citealt{hammer_etal_94,benjam_etal_05,cabrer_etal_08,wegg_etal_15}). Initially the data appeared to show that the in-plane long bar substantially misaligned with the Milky Way's bulge-bar. \citet{mar_ger_11} offered a plausible explanation for a slight misalignment of the two components: the bulge-bar could have developed leading ends through interaction with the adjacent spiral arm heads. A large misalignment was not supported by the new analysis of \citep{wegg_etal_15} that is based on a more homogeneous analysis of a wider range of data, and limits a possible misalignemnt to a few degrees. The aligned bulge-long bar structure found by their work agrees with results from simulations and photometry of barred galaxies, while two misaligned bars with similar sizes are not observed in external galaxies. In any case, two independently rotating bars should align with each other through dynamical coupling in a few rotation periods, unless one bar is much larger than the other \citep{deb_she_07,she_deb_09,du_etal_15}, which is not the case in the Milky Way. We augment the mass in the N-body model long bar with a component described by the following equation:

\begin{eqnarray}\label{eq:lb}
\Phi_{\rm lb}(R,\varphi)=
\begin{array}{ll}
-\zeta_{\rm lb}R&e^{-\frac{R^2}{\sigma^2}[1-\cos(m\phi-f_{\rm m}(R))-R/\epsilon_{\rm lb}]}  \\
~~~~ \times&e^{-(R-R_{\rm lb})^2/2\sigma_{r}^2} e^{-(\varphi-\varphi_{\rm lb})^2/2\sigma_{\varphi}^2}.      
\end{array}
\end{eqnarray}

This is the spiral arm potential tapered by two Gaussian function though $R$ and $\varphi$ direction. For our best model, we adopt $m=2$, $R_i=8\kpc$, $\gamma=139.5^\circ$ and $i=12.5^\circ$, which are the same with the spiral arm part. The amplitude $\zeta_{\rm lb}$, the scale length $\epsilon_{\rm lb}$, and the half-width $\sigma$ are $-2000 \,(\rm km/s)^2$, $3.8\kpc$, and $5.5\kpc$, respectively. For the two Gaussian functions, we use $R_{\rm lb}=5.0\kpc$, $\varphi_{\rm lb}=2.5\degree$, $\sigma_{r}=0.8\kpc$ and $\sigma_{\varphi}=15\degree$, giving a shape leading by $2.5\degree$. Using the method described in Section~\ref{sec:sppotential}, we estimate the mass of the additional long bar component to be $8.2\times10^{9}\Msun$ within $2z_{0}$ $(0.1\kpc)$, which is in good agreement with the value of $7.3-8.8\times10^{9}\Msun$ measured by \citet{wegg_etal_15}. 

\section{Best-fitting Model For The Milky Way}

\label{sec:simulationresults}

We show our best-fitting gas flow model for the Milky Way in Figure \ref{fig:besfit}. This model contains the potential from P15, together with a nuclear bulge, two pairs of $m=2$ spiral arms and a long bar component as described in the last section. The pattern speed of the bar is $33\freq$ and for the spiral arms it is $23\freq$. The gas surface density map is shown in the left panel and the corresponding \lvplot\ is shown in the right panel. To obtain the \lvplot, we assume the Sun is located at $(x,y)=(0\kpc,-8.3\kpc)$ \citep{chatzo_etal_15} with a velocity vector of $(-210\kms,0\kms)$. The bar angle to the Sun-Galactic Center line is $27\degree$, and the bin size in the \lvplots\ is $\Delta l = 0.6\degree$ and $\Delta v = 3\kms$. 

The use of the term `best-fitting' does not mean that we fit the model to the observational data, which is a challenging task (also see \citealt{sor_mag_15}). We use the criterion `approximately reproduces selected features' (i.e. the black lines and dots in the right panel of Figure \ref{fig:besfit}). Gas flow models that can better reproduce more of the selected features are preferred, and the best-fitting model is the one that reproduces the most features.

From the surface density map, we see that the locations the spiral arms agree reasonably well with those traced by the high-mass star forming regions (HMSFRs) in \citet{reid_etal_14} (white dashed lines). The minor offset between the model and the observations is due to our  use of a uniform constant pitch angle ($12.5\degree$) for all of our four spiral arms, whereas the results of \citet{reid_etal_14} imply that the pitch angle for each arm varies from $6.9\degree$ to $19.8\degree$. Nevertheless, considering the thickness of the spirals and the measurement errors in the observations, the model with one common pitch angle seems reasonable to describe the locations of the observed spiral arms at $R \gtrsim7\kpc$. Note that the Local arm (and its counterpart at the opposite side of the Galaxy) here is self-consistently generated by the imposed spiral arm potentials, and it has a slightly small pitch angle compared to the four main spiral arms. For $R\le7\kpc$, the gas flow is mainly dominated by the bar, and the spiral arms in this region align themselves with the bar with a larger pitch angle. In the innermost region ($R\le4\kpc$), we see a typical gas flow driven by the bar, with a pair of off-axis shocks (corresponding to the dust-lanes) and a circum-nuclear ring, which is commonly observed in nearby barred galaxies (e.g., \citealt{san_hun_76,athana_92b,but_com_96,martin_etal_03a}). 

The right panel of Figure \ref{fig:besfit} shows the \lvplot\ of the model. The black lines and symbols depict the features identified by \citet{rod_com_08} and the terminal velocity curves deduced from various observations, respectively. At negative $l$, the tangent points around $l=-50\degree$ and $-70\degree$ are very well reproduced, but around $l=-30\degree$ the terminal velocity of the model is $\sim30\kms$ lower than the observations. At positive $l$, the tangent point around $l=30\degree$ is offset in the model by about $5\degree$, and the tangent point around $l=50\degree$ is not obvious as the Sagittarius arm here is weak. The model seems to produce more features on the \lvplot\ compared to the observed one, as the spiral arms in our model are continuous all the way from the Galactic Center to the outer part, and we have two more weak spiral arms (the Local arm and its counterpart at the other side) generated by the imposed spiral potential. The former leads to some features like the segment between $l=-10\degree$ to $l=-50\degree$ which should be a part of the Scutum arm, and the later leads to some features like the segment at the right side of the Sagittarius arm which should be a part of the Local arm (See Figure \ref{fig:features} for a more clear view).

The Galactic bar mainly dominates for $l\le30\degree$, as the bar ends at $\sim 27\degree$. In the \lvplot\ the Molecular Ring corresponds to the four strong arms at the bar end. The formation of these four arms may be due to the 4:1 resonances. The Near and Far 3-kpc arms are the arms that wrap up the bar, they are roughly reproduced in the \lvplot. The bump at $l\sim10\degree$ is $\sim30\kms$ lower than the observations (some studies call this feature 135-km/s arm, e.g. \citealt{fux_99b}). In the very central region, the Connecting arm and the Bania's Clump 2 are $\sim3\degree$ offset from the observations, these features are associated with the dust-lane shocks. The CMZ is associated with the nuclear ring, which is slightly offset compared to the observations. The peak of the \lvplot\ at $l=3\degree$ is also roughly reproduced, but at $l=-3\degree$ the model is $\sim40\kms$ larger than the observed one. As the gas in our simulations is roughly symmetric about the origin, while the observed dust-lanes and nuclear ring in our Galaxy are slightly lopsided \citep{bally_etal_87,molina_etal_11}. The discrepancy between the model and the observations in the \lvplot\ may due to the lack of lopsidedness in the model.

We also give a better view for the links between the real structures in the face-on image and the features in the \lvplot\ (Figure \ref{fig:features}). We see that the spirals form continuous ridges in the \lvplot, and some of them overlap with each other. The Molecular Ring is mainly dominated by two of the bar-driven spiral arms around the bar end, and by the Scutum arm. The other two bar-driven spiral arms around the bar end form the Near-3kpc arm and the 135-km/s arm, and the Far-3kpc arm is actually some feathers (or a weak spiral arm) between the Molecular Ring and the 135-km/s arm. The Connecting arm and the vertical features correspond to the two dust-lane shocks and the gas clumps on them. The face-on view of the gas flow here looks different from the reference model in \citet{sorman_etal_15c}, although both of the models give a good representation to the \lvplot, suggesting some degeneracies in the $(l,v)$ space. 

Despite minor differences between the model and the observations, the morphology and kinematics are in good agreement with observations, which means a low pattern speed model can also work for our Milky Way, although we need more components than a simple barred potential. We explain why we need the nuclear bulge, the spiral arms, and the long bar component in the next section.

\section{Model Variations}

\label{sec:simulationvariation}

\subsection{The effects of the nuclear bulge}

We first demonstrate why a nuclear bulge is necessary in the center. Figure \ref{fig:nb} illustrates the effect of adding the nuclear bulge. We see that there is an $\xone$-type ring (which is elliptical and elongated along bar major axis, see the definitions in \citealt{kim_etal_12a}) in the left panel by using the potential from P15 only. We know that the $\xone$-type ring is rare in nature and there is no such a feature in our Milky Way. According to \citet{li_etal_15}, decreasing bar pattern speed or increasing bulge central density could turn an $\xone$-type ring into an $\xtwo$-type ring (which is nearly circular and commonly observed). We also see in Figure \ref{fig:rotcurve} that in order to generate a inner Lindblad resonance (ILR) for the potential from P15, the pattern speed of the bar needs to be less than $\sim35\freq$. However, we have tried various bar pattern speeds even down to $10\freq$, the $\xone$-type ring still exists, which implies that changing bar pattern speed alone cannot generate an $\xtwo$-type ring. This is probably due to the positive range of $d(\Omega-\kappa/2)/dR$ at $R<1.2\kpc$, which makes the gas form a pair of leading nuclear spirals at the beginning then quickly turn into an $\xone$-type ring (\citealt{combes_96}). The $d(\Omega-\kappa/2)/dR$ at $R<1.5\kpc$ can be modified to be negative simply by adding a dense center \citep{eng_ger_99,li_etal_15}. Therefore we need more mass in the central region of P15 to generate a reasonable nuclear ring/disk. 

Observations have shown evidence for a dense component in the very central part of the Milky Way. \citet{launha_etal_02} measured the COBE near-IR light at the Galactic Center and found a nuclear bulge/disk of around $300\pc$ in radius and $45\pc$ in height. This component was not included in P15 beacuse they were only able to go to $\sim1\degree$ from the plane before extinction and crowding became too high, while the nuclear bulge/disk becomes significant at $\sim50\pc$ or $\leq0.5\degree$. By assuming a mass-to-light ratio $\gamma_{\odot}$ of 2, the mass of the nuclear bulge is $1.4\times10^{9}\Msun$. We adopt their results and the corresponding gas surface density is shown in the right panel of Figure \ref{fig:nb}. Now a typical bar-driven gas flow pattern appears, with a pair of dust-lanes and an $\xtwo$-type nuclear ring. Note that adding such a nuclear bulge changes only the central region ($R\le1.5\kpc$) of the gas flow; the outer region is nearly the same for these two models. 

The $\xtwo$-type nuclear ring corresponds to the parallelogram-shaped CMZ in the \lvplot, which has been studied for a long time. The size of the nuclear ring is an important parameter to constrain the shape of the potential as the gas in the ring follows $\xtwo$ orbits (\citealt{binney_etal_91}). In the observed \lvplot\ the CMZ spreads from $\sim-1.5\degree$ to $\sim2\degree$ (\citealt{bally_etal_87}), which is similar to the size of the nuclear ring in the SPH simulation done by \citet{Kim_etal_11}. \citet{molina_etal_11} found that the cold gaseous nuclear ring (or disk) in the Galactic Center has a radius of $\sim100\pc$ by using the far-infrared cameras on the Herschel satellite. Very recently, \citet{schonr_etal_15} found a $\sim150\pc$ rotating nuclear disk composed by young stars using APOGEE data, which is probably formed from the gaseous nuclear ring. However, the nuclear ring in our simulation has a radius of $\sim300\pc$. Considering the complicated environment in the Galactic Center, it is possible that our assumptions for the gas might be over-simplified, and/or the nuclear bulge potential derived from the COBE image may not be accurate. Both would affect the radius of the nuclear ring. For example, a magnetic field of equipartition strength with the thermal energy of the gas could make the ring size smaller by a factor of $\sim2$ (\citealt{kim_sto_12}). Also, a lower mass-to-light ratio for the nuclear bulge would generate a smaller ring, but it is less sensitive to the bar pattern speed as long as the parameters of the nuclear bulge are fixed.    

\subsection{The effects of the spirals}

\label{sec:effectsofspirals}

\begin{figure*}
\epsscale{1.2} \plotone{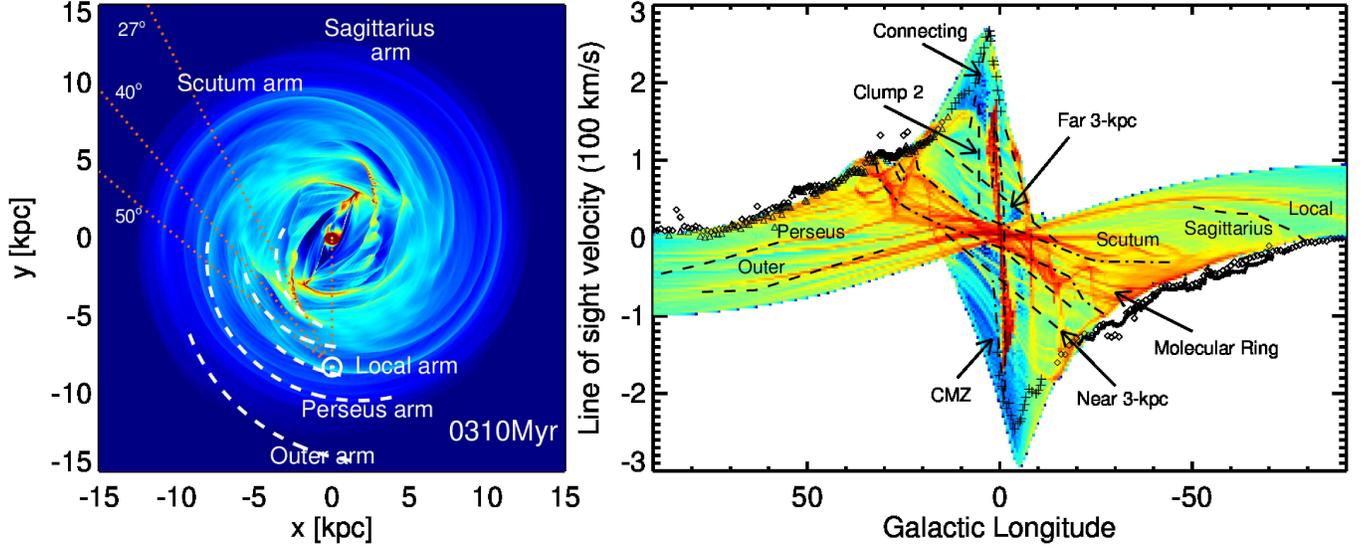} 
\caption{Gas surface density (left panel) of the model without spiral arms and its corresponding \lvplot\ (right panel). All the symbols and lines have the same meaning as in Figure \ref{fig:besfit}.
\label{fig:sp}}
\end{figure*}

\begin{figure*}
\epsscale{1.2} \plotone{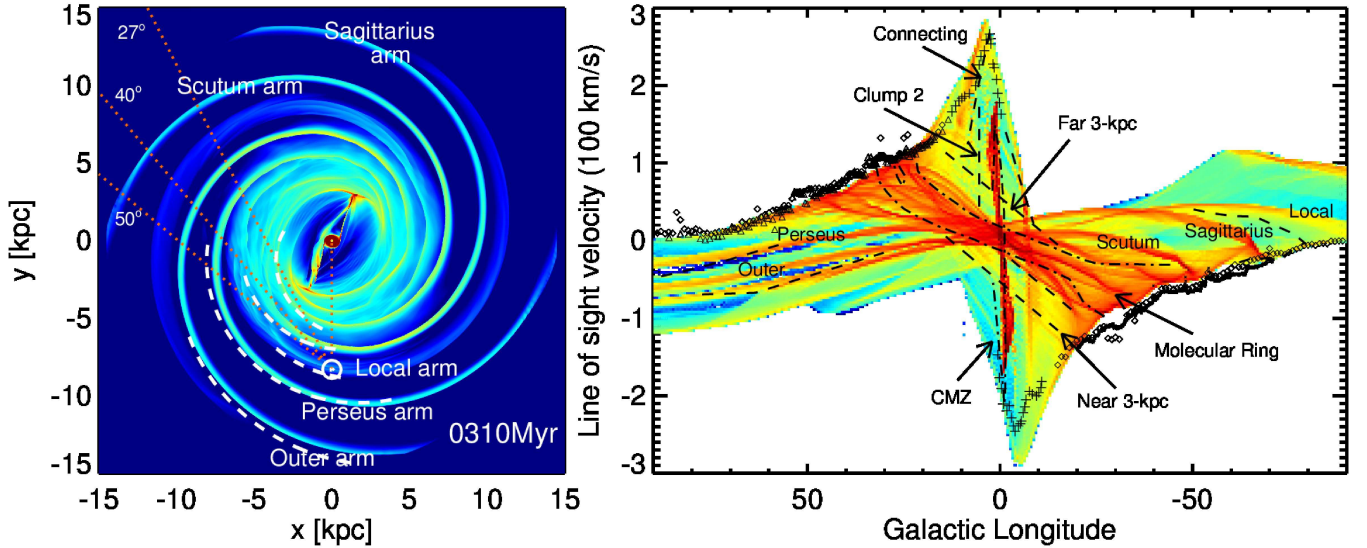} 
\caption{Gas surface density (left panel) of the model without long bar component and its corresponding \lvplot\ (right panel). All the symbols and lines have the same meaning as in Figure \ref{fig:besfit}.
\label{fig:le}}
\end{figure*}

In our best-fitting model we include two pairs of spiral arms with a pattern speed of $23\freq$. Now we remove them from the potentialto isolate their effects. The resulting gas surface density and \lvplot\ are shown in Figure \ref{fig:sp}. Without the imposed spiral arms, the bar drives the gas flow inwards inside the co-rotation radius, as can be seen in the left panel of this figure, but the spiral arms driven by the bar in the outer regions are weak compared to our best model with an imposed spiral potential, and they have a flocculent structure around the bar co-rotation radius ($R_{\rm CR}=6.3\kpc$). Therefore if we aim to reproduce the well-defined, continueous spiral arms suggested by the data of \citet{reid_etal_14} (white dashed lines), we need to impose additionally a spiral arm potential (see also \citealt{bissan_etal_03}).

One may argue that the self-gravity of gas may also help to form spirals without adding an external potential, but the resulting spirals may be transient and sensitive to the density of the initial gas disk. The more plausible scenario may be that the gas responds to the stellar potential as in our best-fitting model. This is supported by a recent study by \citet{hou_han_15}. They reported an obvious offset between the stellar spiral arms and the gas spiral arms in the Milky Way, which means the existence of a quasi-stationary density wave in our Galaxy, and the gas motion is mainly dictated to the distribution of stars. 

In the inner region of the left panel of Figure \ref{fig:sp}, the four strong arms at the bar end are driven by the N-body bulge-bar and the thin long bar potential, therefore they still exist and produce the Molecular Ring in the \lvplot. However, the Connecting arm, the Near and Far 3-kpc arm, and Bania's Clump 2 can barely be identified from the \lvplot. This is because the imposed spiral arm potential with a different pattern speed channels gas flow inwards from the outer regions of the Galaxy, making those features relatively long-lived and prominent. Without the imposed spiral potential, these features are obvious only at the beginning of the simulation but decay in a short time (\citealt{seo_kim_14}), as the gas flows to the center along the shocks and accumulates in the central region, leading to the dissipation of the shock features with time. In the outer region of Figure \ref{fig:sp}, the arms driven by the bar are not obvious, and the tangent points at ($|l|\ge30\degree$) are poorly reproduced. This is due to that the bar perturbation is weak outside the co-rotation radius, and the gas here still follows nearly-circular orbits.

\subsection{The effects of the long bar}

\label{sec:leeffects}

\begin{figure*}
\epsscale{1.2} \plotone{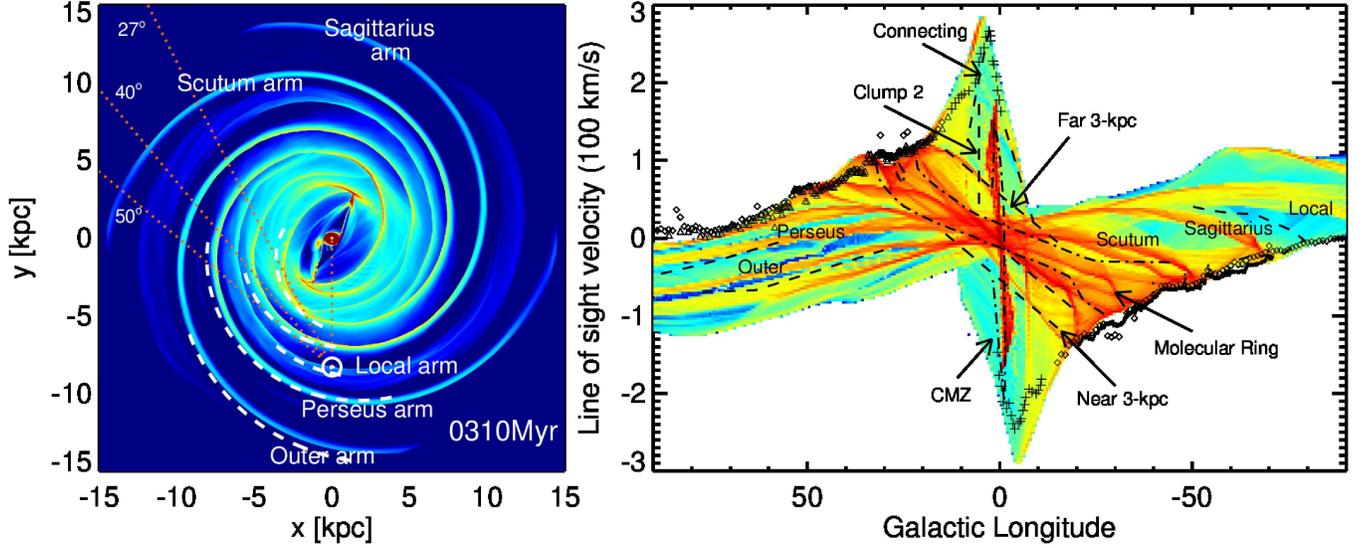} 
\caption{Gas surface density (left panel) of the model with additional force described in Section~\ref{sec:diffrotcurve} together with a slightly more massive long bar, and its corresponding \lvplot\ (right panel). All the symbols and lines have the same meaning as in Figure \ref{fig:besfit}.
\label{fig:lvrotcurve}}
\end{figure*}

We remove the long bar component from our models to isolate the effects of this structure. The resulting gas surface density and \lvplot\ are shown in Figure \ref{fig:le}. As this model includes the spiral arm potential, the tangent points, Connecting arm and Bania's Clump 2 are better reproduced compared to the model in the last section. However, in this model the Molecular Ring seems to cover a larger region, and the bar-driven spiral arms at the bar end are quite weak. The reason why our model without the long bar component has so much less gaseous structure in the region around the end of the bar is because it does not have enough quadrupole potential there (also see \citealt{sorman_etal_15c}). 

The additional mass added to the long bar in our best model generates four strong arms around the bar end, which have a larger pitch angle compared to the four large-scale spiral arms (Figure \ref{fig:besfit} and \ref{fig:sp}). Two of these four arms, connecting to the Scutum arm and the Perseus arm, result in a better description of the Near and Far 3-kpc arms than in the model without the additional mass. Therefore we conclude that the long bar component is important for generating (Near and Far) 3-kpc arms that match the observations well.

The leading twist angle $\varphi_{\rm le}=2.5\degree$ we use in Equation \eqref{eq:lb} is not very important, i.e. if we set $\varphi_{\rm le}=0\degree$, the gas flow pattern is almost the same. The only difference is the existence of gas clumps (e.g., Bania's Clump 2) on the dust-lanes at a given epoch. In the simulation with $\varphi_{\rm le}=0\degree$, the gas clumps disappear at the moment when the spirals have finished a whole rotation relative to the bar ($310\Myr$), although they do appear at earlier times. This certain epoch is when the spiral arms arrive at the right position relative to the Sun, as in the bar co-rotation frame the spiral arms are rotating with respect to the bar. Since it is difficult to control the exact formation time of the transient gas clumps on the dust-lanes, which tend to be stochastic, we still keep this very slight leading twist in our best-fitting model. These gas clumps mainly form the vertical features in the \lvplot\ as discussed in Section \ref{sec:discussion}.

\subsection{The effects of variations in the rotation curve}
\label{sec:diffrotcurve}

The exact rotation curve of the Milky Way is uncertain. In this section we show that this does not influence the results of this work. For example, the rotational velocity measured at the Sun radius is $V_{\rm LSR} = 238\kms$ from \citet{reid_etal_14} which is about $\sim20\%$ larger than $210\kms$ adpoted in our best-fitting model. The difference between the circular velocity at the flat part in our model $V_{flat}\sim210\kms$ outside $\sim2\kpc$ and in \citet{reid_etal_14} $V^{\prime}_{flat}\sim238$ outside $5.5\kpc$ can be described by $\Delta V = V^{\prime}_{flat}-V_{flat} = 28\times(R-2)/(5.5-2)$ which results in terminal velocity differences between $0-28\kms$. But the solar velocity projected with the factor $\rm sin$$(l)$ also increases by $7-19\kms$ ($2\kpc \le R \le 5.5\kpc$ corresponds to $14\degree \le l \le 42\degree$, which leads to $\rm sin$$(l) \times 28\kms \sim 7-19\kms$). So the resulting differences in the \lvplot\ between our best-fitting model and a model that uses the rotation curve of \citet{reid_etal_14} would be at an order of O($10\kms$), which is relatively small compared to the differences in the rotation curves. 

We have also confirmed through simulations that the changes are small in the \lvplot\ by varying the rotation curve. We artificially add into our best-fitting model an extra radial force $F_{\rm extra} = (V^{\prime2}_{flat}-V_{flat}^2)/R = (V^{\prime}_{flat}-V_{flat})(V^{\prime}_{flat}-V_{flat}+2V_{flat})/R$ like:
\begin{eqnarray*}
\label{eq:extra}
F_{\rm extra}(R) =
\left\{\begin{array}{ll}
0  ~~~~~~~~~~~~~~~~~~~~~~~~~~~~~~~~~~~~~~~~~~~~~~~~~~~~~~~~~~~~~~ &R \le 2\kpc \\[0pt]
[28(\frac{R-2}{5.5-2})][(\frac{R-2}{5.5-2})+2V_{flat}]/R  ~~~~ 2\kpc < &R \le 5.5\kpc \\[0pt]
28(28+2V_{flat})/R ~~~~~~~~~~~~~~~~~~~~ 5.5\kpc < &R .
\end{array} \right.
\end{eqnarray*}
The corresponding rotation curve becomes be flat at $R\sim5.5\kpc$ with a flat velocity of $V^{\prime}_{c}=238\kms$. Note that $V_{c}$ of \citet{reid_etal_14} inside $5\kpc$ is nearly unconstrained. We run two experimental models with this potential: one has the same pattern speed of the bar ($33\freq$) and of the spiral arm ($23\freq$) with our best-fitting model, but now the corresponding co-rotation radius is $7.3\kpc$ for the bar and $10.4\kpc$ for the spiral arms; the other has a bar pattern speed of $37.7\freq$ and a spiral pattern speed of $26.3\freq$ to assure  that the corresponding co-rotation radii for the bar and for the spiral arms are nearly the same with our best-fitting model ($R_{\rm CR-bar} =6.4\kpc$ and $R_{\rm CR-spiral}=9.1\kpc$). We find that the model with the same absolute pattern speeds instead of the same co-rotation radii is more similar to our best-fitting model with a relative difference about O($10\kms$). The reason that the pattern speed instead of co-rotation radius matters more is because the bar-related features is well inside $R \le 4\kpc$ ($l \le 30\degree$), and the effective potential here is relatively unchanged if the pattern speeds are fixed. Therefore the  same absolute pattern speeds in this case would give a similar \lvplot\ compared to our best-fitting model. 

To achieve a even better fitting gas flow pattern with the additional force above, the long bar needs to be more massive to get similar stream lines for a larger circular velocity. We then modify the long bar parameters ${\sigma}_{\varphi}$ and ${\zeta}_{lb}$ to be $25\degree$ and $-1200\kms$, respectivly (the values in the best-fitting model are $15\degree$ and $-2000\kms$). The corresponding long bar mass is $8.6 \times 10^9 \Msun$, slightly higher than $8.2 \times 10^9 \Msun$ in the best-fitting model. The resulting gas surface density and \lvplots\ are plotted in Figure \ref{fig:lvrotcurve}. The new model matches some structures in the \lvplot\ better (the far 3-kpc arm, the molecular ring, and the CMZ) and others worse (the near 3-kpc arm, the Connecting Arm, and clumps) for the same pattern speed, and is of similar overall quality. Our conclusion that low pattern speeds are possible is therefore maintained.

According to the experiments above, we think the differences in the rotation curve at $R>2\kpc$ can not significantly change the conclusion of the bar pattern speed predicted in our paper. The mass distribution at $R \le 2\kpc$ is well constrained by the 3D density of red clump stars in \citet{weg_ger_13}, so we do not vary it.

\subsection{Different bar pattern speeds}

\begin{figure}
\epsscale{1.2} \plotone{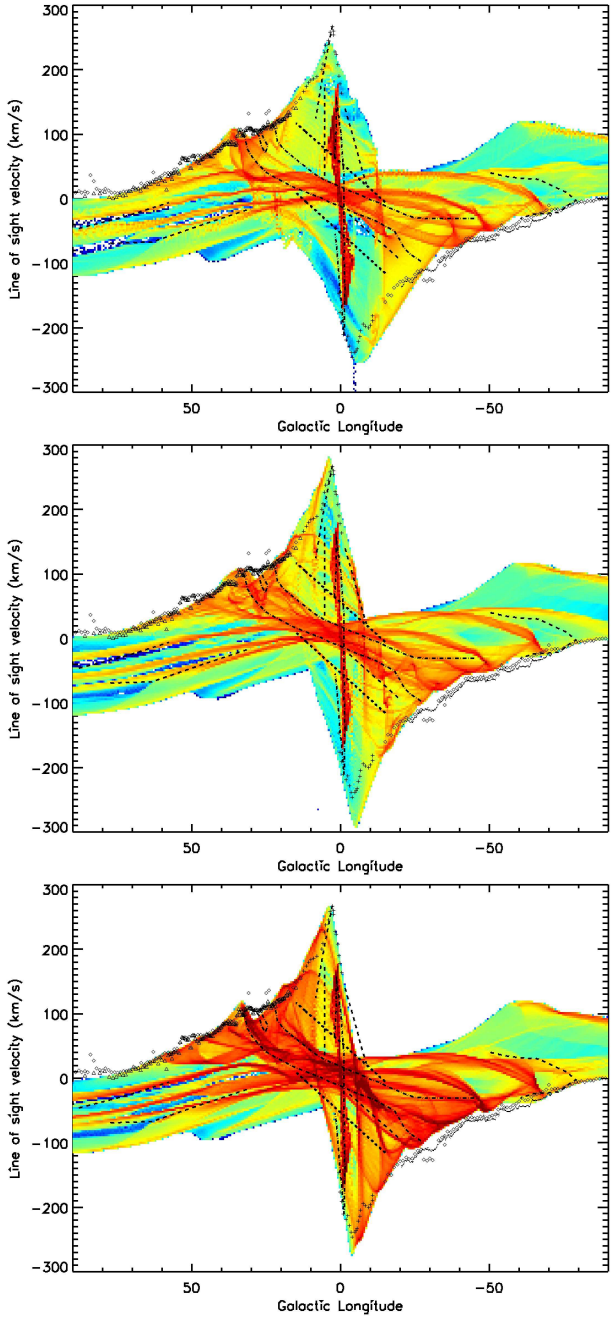} 
\caption{Comparison of models with different bar pattern speeds. All the symbols and lines have the same meaning as in Figure \ref{fig:besfit}. The panels from top to bottom show  models with bar pattern speeds of $23\freq$, $33\freq$ (best-fitting model) and $43\freq$, respectively. The snapshots are taken at a time when the spiral arms have finished a whole rotation relative to the bar. Note that we only change the bar pattern speed; the galactic potential in the corotating frame is the same for all three models.
\label{fig:com}}
\end{figure}

The pattern speed of the bar is an essential parameter to determine the dynamics of the galaxy, but its value in the Milky Way is still under debate. The gas kinematics is often used to constrain the bar pattern speed; thus we vary this parameter in our best-fitting model to see whether other pattern speeds would also produce a reasonable \lvplot.

We run additional two models with bar pattern speed of $23\freq$ and $43\freq$ and plot their \lvplot\ together with our best-fitting model in Figure \ref{fig:com}. We see that a lower or higher bar pattern speed does not reproduce most features in the \lvplot. For the bar pattern speed of $23\freq$ which means a co-rotating bar and spiral arms, the Near 3-kpc arm moves to a lower part, and the tangent point at $l=-30\degree$ becomes $-35\degree$. The Molecular Ring seems to be less prominent, and the forbidden velocities at ($l>0;v<0$) and ($l<0;v>0$) are larger than the observed envelope. Similarly, the model with the bar pattern speed of $43\freq$ gives a steeper 3-kpc arm, moving the the tangent point at $l=-30\degree$ inward to $-20\degree$, and it makes the Molecular Ring extend to a larger region and the forbidden velocity to move below the envelope. We also experimented with other bar pattern speeds within the range $23-43\freq$. Of all the models we tested, the best-fitting model is still that with a pattern speed of $33\freq$.While for our model parametrization the optimal range for the pattern speed is constrained within a few $\freq$, the range may shift somewhat for different potentials in order to maintain the features in the \lvplot. Searching systematically through potential space is, however, challenging, due to the intractably high number of free parameters (\citealt{sorman_etal_15c}). Our main result is therefore that pattern speeds as low as $33\freq$ are consistent with the observed \lvplot.

\section{Discussion}
\label{sec:discussion}

Our models have shown that in order to generate a gas dynamics model that matches the Galactic \lvplot\ well, we need to include a nuclear bulge which helps to generate the nuclear ring; two pairs of spiral arms which continuously channel the gas flow inwards to generate prominent Connecting arm and clumps; and a strong long bar component which generates strong bar-driven arms at the bar end. All these components are motivated by observations, and we can understand how they affect the gas flow reasonably well.

The base potential of this simulation was taken from the M2M model of P15 for the Galactic box/peanut bulge. These authors found a low pattern speed for the bar ($25-30\freq$) by fitting the BRAVA stellar kinematic data in the bulge. Their value is consistent with the result obtained here ($33\freq$). Such a low value is in contrast to the high pattern speeds ($50-60\freq$ obtained in some previous gas dynamical studies \citealt[e.g.,][]{fux_99b,bissan_etal_03,pettit_etal_14}). These high pattern speed models have difficulties in  explaining several features in the \lvplot\ \citep{sor_mag_15}, notably (1) the high-velocity peaks at $l\approx \pm3\degree$; (2) the large forbidden velocities at ($l>0;v<0$) and ($l<0;v>0$); (3) the Near and Far 3-kpc arms; (4) the vertical features, such as Bania's Clump 2. We argue that our best-fitting model has improved in these four aspects:

\subparagraph{(1)} The formation of high-velocity peaks is due to the large velocity jumps at the dust-lane shocks. A lower bar pattern speed or a more massive bar induces a stronger shock \citep{li_etal_15}, which gives higher velocity peaks. These shocks may not have been sufficiently well-resolved in some low-resolution simulations, as argued in \citet{sorman_etal_15a}.

\subparagraph{(2)} The forbidden velocity regions strongly depend on the bar pattern speed, and depend weakly on the strength and length of the bar quadrupole, as well as on the bar angle, as suggested in \citet{sorman_etal_15c}. Our best-fitting bar pattern speed of $33\freq$ is consistent with their estimated range $30-40\freq$ based on forbidden velocity criteria.

\subparagraph{(3)} The Near and Far 3-kpc arms are sensitive to the quadrupole, which is related to our long bar component. While with enough quadrupole moment, the gas still favors a bar pattern speed around $33\freq$ to give a good match to the 3-kpc arms. \citet{sorman_etal_15c} reached a similar conclusion on this point, but they offered a different explanation on the formation of the vertical features in the $(l,v)$ plane. 

\subparagraph{(4)} \citet{sorman_etal_15c} argue that the vertical features are different portions of the two dust-lane shocks, as the shocks show quite a spread in longitude when projecting to the $(l,v)$ plane (see their Figure 5 and our Figure \ref{fig:features}). As the vertical features are quite strong and distinctive from the shocks in the $(l,v)$ plane, we think they are mainly infalling gas clumps on the dust-lane shocks. We observed in our simulation that some gas clumps fall into the CMZ region along the dust-lane shocks, likely due to the wiggle instability \citep{kim_etal_14}. 

In summary, combining our results with those of authors who found bar pattern speeds in the range ($30-40\freq$, \citealt{wei_sel_99,rod_com_08,shen_14,sorman_etal_15c}), we conclude that a gas flow model in a long, strong, and relatively slowly rotating bar potential gives a better description of the \lvplot\ than in high pattern speed models.

\begin{figure}
\epsscale{1.2} \plotone{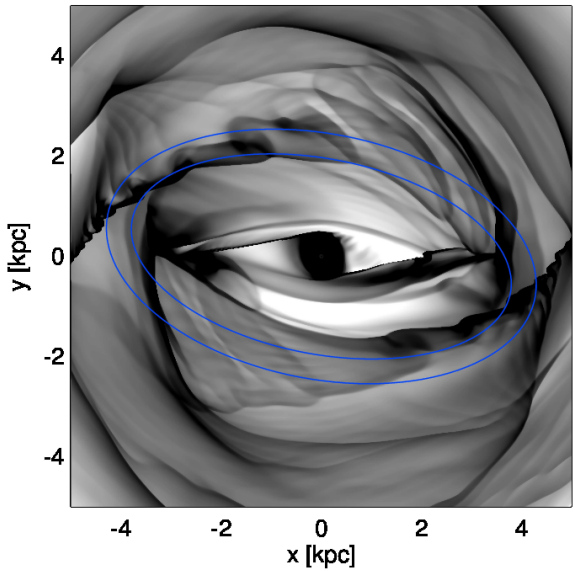} 
\caption{The inner $5\kpc$ of our best-fitting model. The elliptical ring outlines the locus of the 3-kpc arms given in the best-fitting model of  \citet{green_etal_11}, who used 6.7 GHz methanol masers to match the \lvplot. In this plot the bulge-bar lies horizontally. 
\label{fig:3kpc}}
\end{figure}

\citet{dehnen_00} and \citet{antoja_etal_14} suggested a bar pattern speed of $53 \pm 3\freq$ based on the bimodality of the velocity distributions in the solar neighborhood. These authors explained this phenomenon in terms of the orbit shapes near the outer Lindbald resonant (OLR), where our Sun should be located slightly outside the OLR. However, in our best-fitting model the OLR lies far outside the Sun; instead the outer 4:1 resonance is located at $\sim8\kpc$ (Figure \ref{fig:rotcurve}). Whether this could generate similar structures in velocity space remains to be investigated. In addition, \citet{minche_etal_09} and \citet{antoja_etal_09} argued that the influence of the spiral arms on the kinematic structures in the solar vicinity may be as important as that of the Galactic bar.

We note that the short extent of the peanut shape bulge does not necessarily imply a fast-rotating bar, because the peanut shape is not necessarily caused by the vertical inner Lindblad resonance (vILR) as suggested by \citet{pfe_fri_91}. In the models of \citet{portail_etal_15b}, a strong peanut is maintained by families of three-dimensional brezel orbits, while the vILR is present only at radii outside the bulge. The hydrodynamical gas flow models provide an independent measurement of the bar pattern speed, and the value favoured here agrees with the M2M models of P15.

\citet{green_etal_11} used the distribution of 6.7GHz methanol masers to support the presence of a thin long bar with a $45\degree$ orientation from the Sun-Galactic Center line. However, we show in Figure \ref{fig:3kpc} that their data is also roughly consistent with our best-fitting model. The tilted blue ellipse (the 3-kpc arms traced by the masers) which is misaligned with the bulge-bar (horizontal) was thought to be formed by a long bar with a different angle. We see in Figure \ref{fig:3kpc} that the our model can also produce a pair of similar misaligned 3-kpc arms. 

A clear improvement to the current work is to use a more accurate potential, as the parameter space in our best-fitting model is still very large. But such a detailed search for a better model is challenging. The large-scale properties of the bar, the spiral arms, the long bar part, the nuclear component together with the shape of the rotation curve at all radii need to be constrained and improved by further studies. Our assumptions for the gas flow appear reasonable on large-scales, but may be oversimplified in the CMZ region closed to the center. Including more physics in the model might be helpful to explain the asymmetric and tilt properties of the CMZ, although this would also increase the parameter space considerably. 

In summary, we propose a low bar pattern speed gas dynamics model for our Milky Way. We include the nuclear bulge, the spiral arms, and the long bar component to our potential and they are all important to generate related $(l,v)$ features. Our best model can better match the features in the \lvplot\ than previous high bar pattern speed gas models, and we are still working to improve the model.

\acknowledgments

We would like to thank the anonymous referee for providing a constructive report which makes this paper more useful. We appreciate Woong-tae\ Kim and Yonghwi\ Kim for helping us understand the \textit {Athena} code. Hospitality at APCTP during the 7th Korean Astrophysics Workshop is also kindly acknowledged. The research presented here was funded by a grant from the Max-Planck Society under the cooperation agreement with the Chinese Academy of Sciences, and was additionally supported by the 973 Program of China under grant No. 2014CB845700, by the National Natural Science Foundation of China under grants No.11333003, 11322326, 11073037, and by the Strategic Priority Research Program ``The Emergence of Cosmological Structures'' (No. XDB09000000) of Chinese Academy of Sciences. This work made use of the facilities of the Center for High Performance Computing at Shanghai Astronomical Observatory.

\bibliographystyle{apj}

\bibliography{gasdynamics}

\end{document}